\documentclass{iaus}
\usepackage{graphicx}

\title[Massive stars in Wd~1 ] 
{Westerlund~1 as a Template for Massive Star Evolution}

\author[I. Negueruela et al.]   
{Ignacio Negueruela$^1$, 
J. Simon Clark$^2$, Lucy Hadfield$^3$\thanks{Present address: Chester F. Carlson Center for Imaging Science, Rochester Institute of Technology, 54
Lomb Memorial Drive, Rochester NY, 14623, USA }
\and Paul~A. Crowther$^3$ }

\affiliation{$^1$Departamento de F\'{\i}sica, Ingenier\'{\i}a de Sistemas y
  Teor\'{\i}a de la Se\~{n}al, \\Universidad de Alicante, Apdo. 99,
  E03080 Alicante, Spain\\ email: {\tt ignacio@dfists.ua.es} \\[\affilskip]
$^2$ Department of Physics and Astronomy, The Open University,
  Walton Hall, \\Milton Keynes MK7 6AA, United Kingdom \\
$^3$  Department of Physics and Astronomy, University of
Sheffield,\\ Sheffield, S3 7RH, United Kingdom }
\pubyear{2008}
\volume{xxx}  
\pagerange{1--6}
\jname{Title of your IAU Symposium}
\editors{A.C. Editor, B.D. Editor \& C.E. Editor, eds.}
\begin{document}

\maketitle

\begin{abstract}With a dynamical mass $M_{\rm dyn}\sim
  1.3\times10^{5}\:M_{\odot}$ and a lower limit
  $M_{\rm cl}>5\times10^{4}\:M_{\odot}$ from star counts, Westerlund 1 is the 
most massive young open cluster known in the Galaxy and thus the
perfect laboratory to study massive 
star evolution. We have developed a comprehensive spectral
classification scheme for supergiants based on 
features in the 6000\,--\,9000\AA\ range, which allows us to identify
$>30$ very luminous supergiants in Westerlund 1 and $\sim 100$ other
less evolved massive stars, which join the large population of
Wolf-Rayet stars already known. Though detailed studies of these stars
are still pending, preliminary rough estimates suggest that the stars
we see are evolving to the red part of the HR diagram at approximately
constant luminosity.
\end{abstract}

\firstsection 
\section{Introduction}

 As they evolve towards the Wolf-Rayet (WR) phase, massive stars
must shed most of their outer layers. Models
predict that massive stars will become supergiants (SGs) and evolve
redwards at approximately 
constant $L_{{\rm bol}}$, but observations reveal
  a complex zoo of transitional objects, comprising Blue SGs, Red SGs,
  Yellow Hypergiants (YHGs), Luminous Blue Variables (LBVs) and
  OBfpe/WNVL stars, whose 
  identification with any particular evolutionary phase is
  difficult. Understanding this evolution is, however, crucial because 
  the mass loss during this phase completely determines the
  contribution that the star will make to the chemistry of the ISM and
  even the sort of post-supernova remnant it will leave.

Unfortunately, massive stars are scarce and, as this phase is very short on
evolutionary terms, examples of massive stars in transition are
rare. For most of them, 
distances are unknown and so luminosities are known at best to
order-of-magnitude accuracy, resulting in huge uncertainties in other parameters ($M_{*}$, $R_{*}$). The difficulty to place these objects in
the evolutionary sequence is obvious.

The open cluster Westerlund~1 (Wd~1) offers an unprecedented opportunity to
improve this situation. It contains a large population of evolved
stars which have formed at the same time, are at the same distance and
have the same chemical composition. With an age of 4\,--\,5~Myr, Wd~1
contains a rich population of stars in transitional states that can be
used to constrain evolutionary models.

\begin{figure}[ht]
\begin{center}
 \resizebox{\textwidth}{!}{\includegraphics[angle=-90]{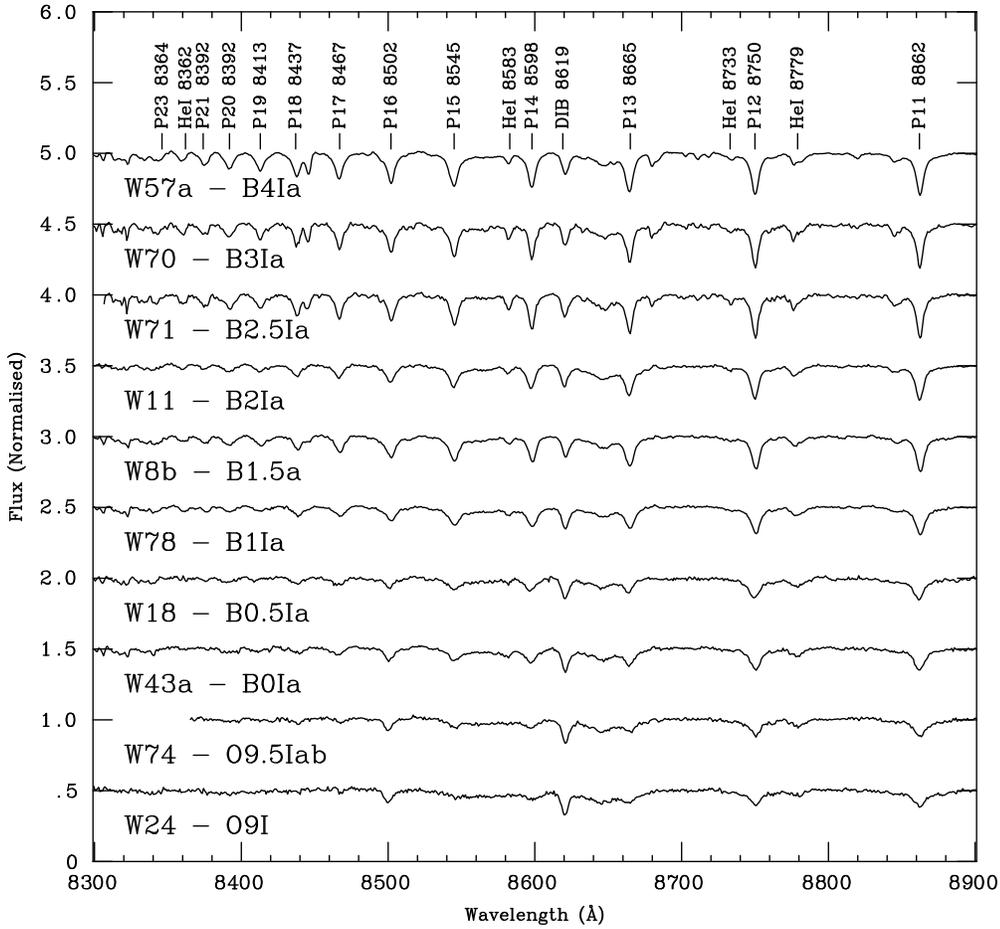}} 
 \caption{Sequence of $I$-band spectra of blue supergiants in Wd~1,
   showing the evolution of the main features. Note the disappearance
   of N\,{\sc i} features between Pa~12 and Pa~13 and O\,{\sc
     i}~8446\AA\ around B2 and the 
   development of Pa~16, which shows the presence of a strong C\,{\sc
     iii} line for stars B0 and earlier..}
   \label{fig1}
\end{center}
\end{figure}

\section{Cluster parameters}

The parameters of Wd~1 are still poorly determined, though significant
progress has been made in recent years. Two estimates of its mass have
been recently made. \cite{mt07} have used the radial velocity
dispersion of the ten stars brightest in the infrared
($\sigma=8.4\:{\rm km}\,{\rm s}^{-1}$) to estimate a
mass of  $\sim 1.3\times10^{5}\:M_{\odot}$. This estimate suffers from
a large uncertainty due to the low number of stars used and the spectral
peculiarities of some. \cite[Brandner et al. (2008)]{brandner} have used
star counts in the 
infrared to set a lower limit on the cluster mass  $M_{\rm
  cl}>5\times10^{4}\:M_{\odot}$. Again, 
there are important uncertainties coming into this estimation. Large
discrepancies have been found in the determinations of distance and
(hence) masses based on infrared pre-main-sequence tracks and optical
post-main-sequence tracks. In any case, the integrated initial mass
function appears consistent with Salpeter's and then the observed
population of massive stars would imply a mass  $\sim
10^{5}\:M_{\odot}$, consistent with both determination.

An important source of uncertainty in both measurements is the role of
binarity, which affects both radial velocities and transformation of
magnitudes into masses. Recent results suggest that the fraction of
binaries with similar mass components is very high amongst cluster
members. Most Wolf-Rayet stars show indirect indications of binarity,
presence of dust in WC stars \cite{crowther} and hard X-ray emission
in WN stars \cite{clark08}. Some of them also show photometric
variability and at least one is an eclipsing binary \cite{bonanos}.
The identification of a large number of X-ray sources detected by {\it
  Chandra} with a population of evolved late-O stars \cite{clark08}
suggests that the high binary fraction extends to lower masses.

The distance to the cluster is not very well constrained
either. Photometry is affected by the very strong reddening
($A_{V}\approx12\:{\rm mag}$). Analysis of the $E(B-V)$ colours
suggests that the 
reddening deviates from the standard law \cite{clark05}. A recent
determination, making use of atomic hydrogen in the direction to the
cluster, gives $d=3.9\pm0.7$ \cite{kothes}, compatible with, though slightly
shorter than, estimates based on the stellar population
\cite[(e.g., Crowther et al. 2006)]{crowther}.

The age of the cluster can be constrained from the observed
population. The ratio of WR stars to red and yellow
hypergiants favours an age of 4.5 or 5.0~Myr, with the progenitors of
the Wolf-Rayet stars having initial masses in the
40\,--\,$55\:M_{\odot}$ range \cite{crowther}. Such age is fully
compatible with the 
observed population of blue supergiants (see below), which should be
descended from stars with initial masses of $\sim 35\:M_{\odot}$. This
again would imply masses of $\leq30\,M_{\odot}$ for the stars at the
top of the main sequence, in good agreement with the O7\,--\,8\,V
spectral type, again appropriate for the age.

\section{Observations and analysis}

 Observations of stars in Wd~1 were carried out on the nights of 2003
 June 12th and 13th using the spectro-imager FORS2 on Unit~1 of the
 VLT (Antu) using three different modes: longslit, multi-object
 spectroscopy with masks (MXU) and multi-object spectroscopy with
 movable slitlets (MOS). We used grisms G1200R and G1028z to obtain
 intermediate resolution spectroscopy. With this setup, we obtained
 almost continuous coverage over the 5800\,--\,9500\AA\ range at
 intermediate resolution.

 Within the central
 $5^{\prime}\times5^{\prime}$ field of view, we selected our targets
 from the list of likely members of \cite[Clark et
   al. (2005)]{clark05}. For the external 
 regions, targets were selected at random amongst relatively bright
 stars. In total, we took three MXU and one MOS mask with both G1200R and
 G1028z grisms, and two further MXU masks with only the G1200R (these were
 aimed at relatively faint objects, which were expected to be OB stars
 near the MS and so not to have strong features in the range covered
 by G1028z). This resulted in $\sim 100$ stars observed with G1200R
 and $\sim70$
 stars observed with G1028z. More than 90\% of the spectra turned out
 to correspond to OB 
 stars, and hence cluster members.

We have used these observations to derive spectral types for cluster
members. The use of $I$-band spectra to classify OB stars has been
explored by \cite[Caron et al. (2003)]{caron} and is further discussed
in Appendix A of 
\cite[Clark et al. (2005)]{clark05}. We have studied further spectral
type and luminosity 
indicators in the $R$ and $I$ bands and derived spectral types from a
combination of features. Features that contribute to our analysis are:
the shape and strength of emission (P-Cygni profiles) in  
H$\alpha$, the strength of the C\,{\sc ii}~6578,6582\AA\ doublet (in
absorption), the presence and strength of the C\,{\sc ii}~7231,
7236\AA\ doublet (in emission, a wind feature), the strength of the
O\,{\sc i}~7774\AA\ 
triplet and the strength of the N\,{\sc ii}~6482\AA\ line. In the $I$
band, apart from the shape and strength of the Paschen lines (see
\cite[Clark et al. 2005]{clark05}), a main indicator is the presence of the
C\,{\sc iii}~8502\AA\ line, which appears, blended into Pa~16, for
stars B0 and earlier (see Fig.~\ref{fig1}).

\begin{figure}[ht]
\begin{center}
 \resizebox{\textwidth}{!}{\includegraphics[angle=-90]{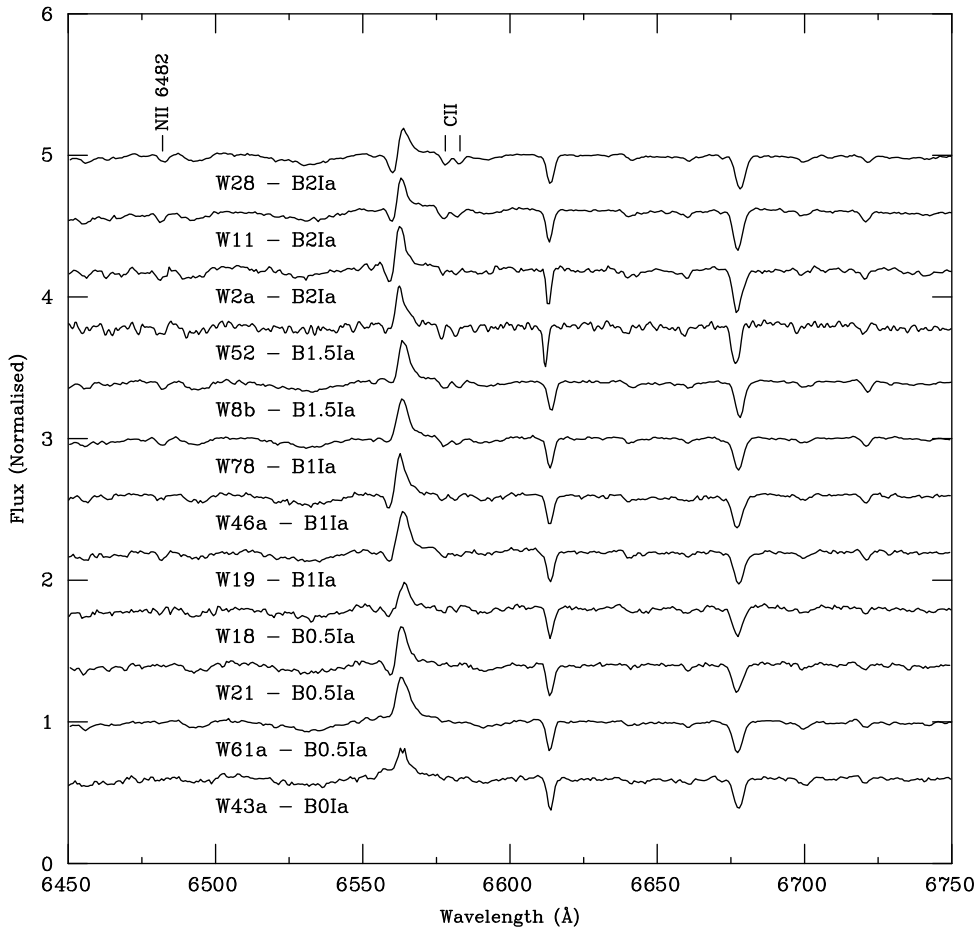}} 
 \caption{Sequence of red spectra of blue supergiants in Wd~1. The
   shape of H$\alpha$ changes from a P-Cygni profile to pure emission
   around spectral type B0.5. The N\,{\sc ii}~6482\AA\ line is only
   prominent in Ia supergiants and its strength peaks sharply at B2. }
   \label{fig2}
\end{center}
\end{figure}

\section{Implications}

{\underline{\it The upper HR diagram}}. The stellar content revealed
by these observations is as follows. There is an A supergiant showing
spectral variability (W243; \cite[Clark \& Negueruela 2004]{cn04}) and
six other A/F 
hypergiants of very high luminosity, with $M_{V}\approx -9$
to $-10$ \cite{clark05}. We call these objects hypergiants because of
their very high intrinsic magnitudes, without making any direct
inference of their evolutionary status. 

The same can be said of at least three late-B luminous stars: W42a
(B8\,Ia$^{+}$), W33 (B5\,Ia$^{+}$) and W7 (B5\,Ia$^{+}$). In addition,
there are $\sim 20$ supergiants that can be unambiguously classified
as Ia, covering the B0\,Ia\,--\,B4\,Ia range. The earliest very luminous
star that we find is W74, with spectral type O9.5 and a luminosity
class that could be Ia or Iab. 

Even though the exact luminosity class of O-type supergiants is more
difficult to determine using red spectra, we do not find evidence for
any other luminous O-type supergiant, though there are $>50$ and
(taking into account the incompleteness of our dataset) likely $\sim 100$
stars which we generically classify as O9\,I, meaning late-O stars
well above the main sequence.

\begin{figure}[ht]
\begin{center}
 \resizebox{\textwidth}{!}{\includegraphics[angle=-90]{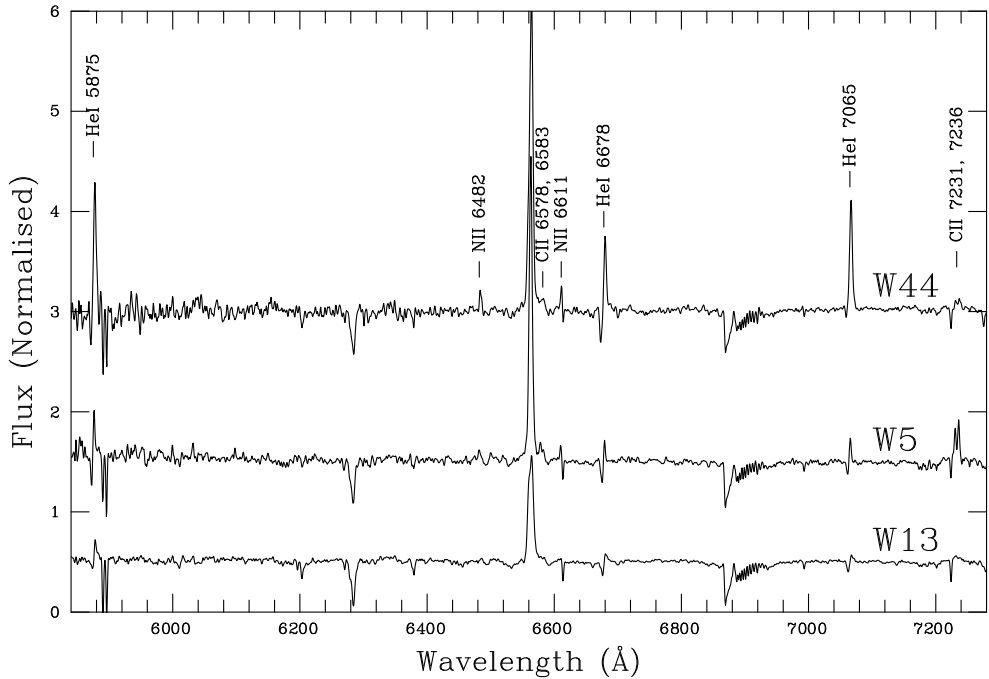}} 
 \caption{Spectra of three transitional objects in Westerlund~1, which
   can be classified as very late WN stars or early-B
   hypergiants. These objects might be examples of stars looping back
   to the blue region from the red supergiant phase.}
   \label{fig3}
\end{center}
\end{figure}

{\underline{\it Evolution}}. We have made a rough attempt at
calculating the intrinsic luminosities of all the supergiants, by
using their $V$ and $I$ magnitudes and assuming that the relation
between $E(V-I)$ and $M_{I}$ is standard. We use the intrinsic
$(V-I)$ colours of \cite{ducati} and bolometric corrections from
\cite{martins} and \cite{humphreys}. Even though there is a large
scatter, we find no clear trend. Stars of all spectral types seem to
have similar bolometric magnitudes, a result consistent with the idea
of redwards evolution at constant bolometric luminosity. Note that we
have not compared the bolometric magnitudes of the red supergiants, as
their spectral types are uncertain, and the bolometric corrections of
such bright red stars are unknown.

This result does not imply that there are no stars looping back
towards the blue in Wd~1. Obviously, the WR stars must have come back
to the blue before reaching their present stage. There are three
emission-line stars in the cluster which can be classified as very
late WN stars or extreme B supergiants (see Fig.~\ref{fig3}). These
objects might be evolving into Wolf-Rayet stars. Detailed analysis of
their spectra, in order to derive chemical abundances, will be
necessary before their evolutionary status may be ascertained. Similar
analyses should be conducted for the A and F hypergiants and W243. 

The next steps in our work involve an investigation of the binary
fraction in the 
cluster and detailed analysis of the spectra with state-of-the art
models in order to obtain a better constraint on their evolutionary
status.

\begin{discussion}

\discuss{Lang}{What does the distribution of radio emission reveal in
  this cluster -- both the extended emission and the point-like sources?}

\discuss{Negueruela}{There are point sources associated with most of
  the cool and cold supergiants, which, we believe, are due to
  photoionisation of their extended atmospheres or winds by the hot
  stars. There is no extended emission obviously associated with the
  cluster, suggesting that supernova explosions have blown away any
  diffuse material. A more detailed summary may be found in the
  contribution by Sean Dougherty and Simon Clark to the proceedings of
  {\it Massive Stars: Fundamental Parameters and Circumstellar
    Interactions}, {\tt arXiv:0705.0971v1}. }

\discuss{Crowther}{As your title suggests, the presence of large
  numbers of massive stars in Wd~1 allows robust tests of evolutionary
models. By way of example, N(WN with H) $<$ N(WC) $<$ N(WN without H),
yet both single star (i.e., Geneva) and binary models (e.g., Eldridge,
this conference) predict the completely wrong subtype distribution of
N(WN without H) $<$ N(WN with H) $<$ N(WC) for an instantaneous burst
of 4\,--\,5 Myr.}

\discuss{Negueruela}{Indeed. The observed population is most likely
  compatible with a single burst of star formation and hence offers an
observational test of evolutionary models. Also, as I mentioned, the
fact that we can observe a large population of massive
($M_{*}>30\,M_{\odot}$) stars evolving away from the main sequence and
a very large population of low mass stars 
evolving towards the main sequence gives us an excellent opportunity to
try to set post and pre-main-sequence isochrones in the same reference
time and compare their predictions.}

\end{discussion}

\end{document}